\documentclass[prd,twocolumn,showpacs,amsmath,amssymb,superscriptaddress,floatfix,nofootinbib]{revtex4}
\usepackage{mathrsfs,bm}
\usepackage{longtable,lscape}
\usepackage{txfonts}
\usepackage{amssymb}
\usepackage{indentfirst}
\usepackage{graphicx,,booktabs}
\usepackage{multirow}
\usepackage{color}
\usepackage{amssymb}
\definecolor{cover}{rgb}{0.77,0.87,0.88}
\definecolor{blueone}{rgb}{0.1,0.1,.7}
\definecolor{citec}{rgb}{0.14,0.47,0.09}
\definecolor{two}{rgb}{0.0,0.5,0.}
\definecolor{three}{rgb}{.5,.1,0.15}
\usepackage[bookmarks=true,bookmarksopen=false,plainpages=false,breaklinks=true,
   bookmarksnumbered=true,hypertexnames=false,
   filecolor=blue,urlcolor=three,menucolor=three,
   linkcolor=three,citecolor=blueone, colorlinks,
   anchorcolor=blue,runcolor=pink,frenchlinks=red
   pdfstartview=FitH,pdftitle=title,%
   pdfauthor=author]{hyperref}

\usepackage{relsize}
\usepackage{xspace}
\def\babar{\mbox{\slshape B\kern-0.1em{\smaller A}\kern-0.1em
    B\kern-0.1em{\smaller A\kern-0.2em R}}}

\begin{document}
\title{Production of  charmed baryon $\Lambda_c(2940)$ by kaon-induced reaction on a proton target}

\author{Yin Huang}
\email{huangy2014@lzu.cn} \affiliation{Research Center for Hadron and CSR Physics, Lanzhou University and Institute
of Modern Physics of CAS, Lanzhou 730000,China}
\affiliation{School of Nuclear Science and Technology,
Lanzhou University, Lanzhou 730000, China}
\affiliation{Institute of Modern Physics,
Chinese Academy of Sciences, Lanzhou 730000, China}
\author{Jun He \footnote{corresponding author}}
\email{junhe@impcas.ac.cn}
\affiliation{Research Center for Hadron and CSR Physics, Lanzhou University and Institute
of Modern Physics of CAS, Lanzhou 730000,China}
\affiliation{Institute of Modern Physics,
Chinese Academy of Sciences, Lanzhou 730000, China}
\author{Ju-Jun Xie }
\email{xiejujun@impcas.ac.cn}
\affiliation{Research Center for Hadron and CSR Physics, Lanzhou University and Institute
of Modern Physics of CAS, Lanzhou 730000,China}
 \affiliation{Institute of Modern Physics,
Chinese Academy of Sciences, Lanzhou 730000, China}
\author{Xurong Chen}
\affiliation{Research Center for Hadron and CSR Physics, Lanzhou University and Institute
of Modern Physics of CAS, Lanzhou 730000,China}
 \affiliation{Institute of Modern Physics, Chinese Academy of Sciences, Lanzhou 730000, China}
\author{Hong-Fei Zhang}
  \affiliation{School of Nuclear
Science and Technology, Lanzhou University, Lanzhou 730000, China}\affiliation{Institute of Modern Physics, Chinese Academy of
Sciences, Lanzhou 730000, China}

\date{\today}
\begin{abstract}
We investigate the possibility to study the charmed baryon $\Lambda_c(2940)$
by kaon-induced reaction on a proton target.  By assuming the $\Lambda_c(2940)$ as a
$pD^{*0}$ molecular state with spin-parity $J^{p}=1/2^{\pm}$, an effective
Lagrangian approach  was adopted to calculate the cross section, the $D^0p$ invariant mass spectrum  and Dalitz plot of the $\Lambda(2940)$ production.
The total cross section of the $K^{-}p\to\Lambda_c(2940)D_s^{-}$ reaction
is found at an order of magnitude about 10 $\mu$b.  By considering the subsequential
decay $\Lambda_c(2940)\to{}D^0p$ with contributions from the $\Lambda_c(2286)$ and the $\Sigma_c(2455)$
as background,  the $K^{-}p\to{}D_{s}^{-}D^{0}p$
reaction are studied.  It is found that the $\Lambda_c(2940)$ is produced mainly at forward angles. The $\Lambda_c(2940)$ signal is predicted to be significant in the $D^0p$ invariant mass spectrum and the Dalitz plot of the $K^{-}p\to{}D_{s}^{-}D^{0}p$ reaction.  The results suggest that it is promising to study the $\Lambda_c(2940)$ with high-energy kaon beam on a proton target in experiment.
\end{abstract}

\pacs{13.60.Le, 12.39.Mk,13.25.Jx}

\maketitle
\section{INTRODUCTION}

Thanks to the experimental progress,  more and more charmed
baryons have been observed. The charmed baryon $\Lambda_c(2940)$  was first observed by
the \babar~Collaboration~\cite{Aubert:2006sp} in the  $D^0p$ invariant mass spectrum. Later
the Belle collaboration~\cite{Abe:2006rz} also reported the observation of  this state in the
$\Sigma^{0,++}_c(2455)\pi^{\pm}$ invariant mass spectrum.  The mass and width of the $\Lambda_c(2940)$
state reported by both
collaborations~\cite{Aubert:2006sp,Abe:2006rz} are consistent
with each other:

\begin{center}
\begin{tabular}{rcl}
\babar:  &M&=2939.8$\pm$1.3$\pm$1.0 ~ {MeV}\nonumber,\\
        &$\Gamma$&=17.5$\pm$5.2$\pm$5.9   ~{MeV};\nonumber\\
Belle:  &M&=2938.0$\pm1.3^{+2.0}_{-4.0}$  ~ {MeV},\nonumber\\
        &$\Gamma$&=$13^{+8+27}_{-5-7}$ ~ {MeV}.\nonumber\\
\end{tabular}
\end{center}

Since the mass of the $\Lambda_c(2940)$ is just a few MeV below  the
$D^{*0}p$ threhold, it was proposed that this state is an $S$-wave $D^{*0}p$ molecular state with spin parity $J^p=1/2^{-}$, and the obtained decay behavior of the $\Lambda_c(2940)$ is
consistent with the experiment~\cite{He:2006is}.   Later, a study about the strong  and
radiative decays of the $\Lambda_c(2940)$ was performed by Dong and his collaborators~\cite{Dong:2009tg,Dong:2010xv}. Their results indicate
that the $\Lambda_c(2940)$ should be assigned as a $D^{*0}p$ molecular
state with $J^p=1/2^{+}$. In Ref.~\cite{He:2010zq}, with a dynamical study of the $D^{*0}p$ interaction in the
one-boson-exchange model, bound states solutions were found with quantum numbers $0(1/2^{+})$ and $0(3/2^{-})$,
which correspond to isoscalar $S$-wave and isoscalar $P$-wave $D^{*0}p$ molecular state, respectively.   Besides the $pD^{*0}$ molecular
state interpretations of the $\Lambda(2940)$, the possibility to assign it as a conventional charmed baryon
was also discussed in many approaches, such as in the potential model~\cite{Capstick:1986bm},
in the chiral perturbation theory~\cite{Cheng:2006dk}, in the $^3P_0$ model~\cite{Chen:2007xf},
in the relativistic quark-diquark model~\cite{Ebert:2007nw}, in the chiral quark
model~\cite{Zhong:2007gp}, in the Faddeev method~\cite{Valcarce:2008dr}, and
in the mass load flux tube model~\cite{Chen:2009dr}.

Since the internal structure of the $\Lambda(2940)$ is not well understood until now, it will be very helpful if we can obtain more information about the $\Lambda(2940)$ in more experiments.
The existing knowledge about the properties of the $\Lambda_c(2940)$ was obtained from the $e^{+}e^{-}$
collision ~\cite{Aubert:2006sp,Abe:2006rz}. Thus, it will be helpful to understand the $\Lambda(2940)$ if we can observe
it in other production processes. In Refs.~\cite{He:2011jp,Dong:2014ksa},
a proposal was made to study the $\Lambda(2940)$ in the $\bar{p}p$
annihilation which can be performed in the future PANDA detector at FAIR. The production of the $\Lambda_c(2940)$ via a pion-induced reaction on a nucleon target was discussed in Ref.~\cite{Xie:2015zga}. The study of the $\Lambda_c(2940)$ with electromagnetic probe
was also proposed in the $\gamma{}n\to\Lambda_c(2940)D^{-}$ reaction~\cite{Wang:2015rda}.

The high-energy kaon beam is available at OKA@U-70~\cite{Obraztsov:2016lhp} and SPS@CERN~\cite{Velghe:2016jjw}, which provides another opportunity to study the charmed baryon. The kaon beam at J-PARC can be also upgraded to the energy region required in the charmed baryon production~\cite{Nagae:2008zz}.  It is interesting to make a theoretical prediction about the charmed baryon production with the kaon beam.
With an charged kaon beam, the $\Lambda_c(2940)$ can be produced with a proton target, i.e., $K^{-} p \to D^{-}_s\Lambda_c(2940)$ reaction. In such reaction, the $s$-channel is usually suppressed seriously because of very large total energy~\cite{Xie:2015zga, He:2012ud,Huang:2013mua}. The $u$-channel contribution is usually suppressed also and more important at backward angles while the $\Lambda(2940)$ is produced at forward angles through $t$-channel~\cite{He:2013ksa}. Moreover, compared with the pion-induced $\Lambda(2940)$ production, an additional $s\bar{s}$ quark pair creation is needed in the kaon-induced production, so the $u$-channel will be further suppressed. Hence, this reaction should be dominant with the Born terms through $t-$channel $D^{*0}$ exchange, which makes the background very small. In this work, we will study the $\Lambda_c(2940)$
production in the $K^{-} p \to D^{-}_s\Lambda_c(2940)$ reaction
in an effective Lagrangian approach. The Dalitz plot and invariant mass spectrum the for the subsequential decay of the $\Lambda_c(2940)$ in the $K^{-} p \to D^{-}_s\Lambda_c(2940)\to D_s^-(D^0p)$ reaction will be studied also.

This paper is organized as follows.  After the introduction, we will present the
effective Lagrangian and the corresponding coupling constant used in this work.
 In Sec. III, the formulism and the numerical result of the kaon-induced $\Lambda_c(2940)$ (we abbreviate it as $\Lambda_c^*$ hereafter) production
on proton target will be given.
In Sec. IV, the Dalitz plot and invariant mass spectrum of the  $K^{-} p\to{}D_s^{-}D^0p$ reaction will be presented. Finally, the paper ends with the discussion and conclusion.

\section{Effective Lagrangian}

\begin{figure}[h!]
\begin{center}
\includegraphics[bb=100 560 450 720, clip, scale=0.5]{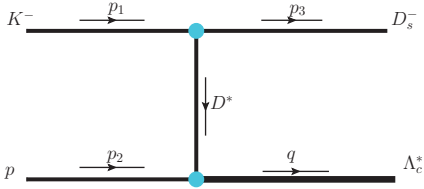}
\end{center}
\caption{(color online). The Feynman diagram for the mechanism of the $\Lambda^{*+}_c$
production in the $K^{-}p\to{}D_s^{-}\Lambda_c^{*+}$
reaction.  We also show the definition of the kinematical ($p_1$,~
$p_2$, ~$p_3$, and~$q$) used in the calculation.  }\label{feydiagrams}
\end{figure}

In Fig.~\ref{feydiagrams}, we illustrate the Feynman diagram for the $t$-channel $K^{-}p\to{}D_s^{-}\Lambda_c^{*}$ interaction, which is the dominant mechanism of the $\Lambda_c^*$ production.
The kaon beam is adopted to attack the proton target, and the $\Lambda_c^*$ is produced by exchange of a $D^*$ meson.  As discussed in the introduction, other production mechanisms will be suppressed heavily and not considered in this work.

\begin{figure}[h!]
\begin{center}
\includegraphics[bb=100 520 450 720, clip, scale=0.5]{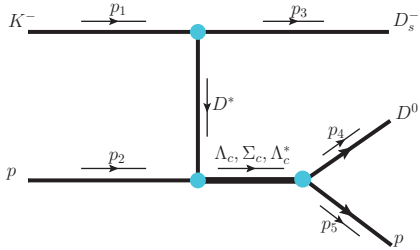}
\end{center}
\caption{(color online). The Feynman diagram for the $K^{-}p\to{}D_sD^{0}p$ reaction. We also show the definition of the kinematical ($p_1$,~
$p_2$, ~$p_3$, ~$p_4$, and~$p_5$)  used in the calculation.  }\label{feydiagrams23}
\end{figure}

To observe the $\Lambda_c^*$, the subsequential decay to $D^0p$ will also be considered as shown in Fig.~\ref{feydiagrams23}. The  $\Lambda_c(2286)^{+}$
and $\Sigma_c(2455)^{+}$ can be produced from the $K^-$ and proton interaction by exchanging a $D^*$ meson and decay to $D^0p$ also. So, in this work, $\Lambda_c(2286)^{+}$
and $\Sigma_c(2455)^{+}$ will be taken as the background of the $\Lambda_c^*$ production.

To compute the amplitudes of the diagrams shown in Figs.~\ref{feydiagrams} and Fig.~\ref{feydiagrams23}, we need the
effective Lagrangian densities for the relevant interaction vertices.
The spin-parity of the $\Lambda_c^{*}$ state was still not determined in experiment.
The theoretical studies~\cite{He:2006is,Dong:2009tg,Dong:2010xv,Dong:2014ksa} suggested that the possible assignment of spin parity of the $\Lambda_c^{*}$
are $J^p=1/2^{+}$ and $1/2^-$. In this work, we will consider these two assignments.
For the $\Lambda^{*}_cpD$ and $\Lambda^{*}_cpD^{*}$ couplings, we
take the Lagrangian densities as used in Ref.~\cite{Dong:2014ksa},
\begin{align}
&{\cal{L}}_{\Lambda_c^{*}pD}=ig_{\Lambda_c^{*}pD}\bar{\Lambda}_c^{*}\gamma_{5}pD^{0}+{\rm H.c.},\\
&{\cal{L}}_{\Lambda_c^{*}pD^{*}}=g_{\Lambda_c^{*}pD^{*}}\bar{\Lambda}_c^{*}\gamma^{\mu}pD^{*0}_{\mu}+{\rm H.c.},
\end{align}
for the assignment $J^{p}=1/2^{+}$, and
\begin{align}
&{\cal{L}}_{\Lambda_c^{*}pD}=g_{\Lambda_c^{*}pD}\bar{\Lambda}_c^{*}pD^{0}+{\rm H.c.},\\
&{\cal{L}}_{\Lambda_c^{*}pD^{*}}=-g_{\Lambda_c^{*}pD^{*}}\bar{\Lambda}_c^{*}\gamma_5\gamma^{\mu}pD^{*0}_{\mu}+{\rm H.c.},
\end{align}
for the assignment $J^{p}=1/2^{-}$.
The coupling constants in the above Lagrangians were determined in Refs.~\cite{Dong:2009tg,
Dong:2010xv} in a hadronic molecular picture with $g_{\Lambda_c^{*}pD}$=-0.54,
$g_{\Lambda_c^{*}pD^{*}}$=6.64 for $J^p=1/2^{-}$ and
$f_{\Lambda_c^{*}pD}$=-0.97, $f_{\Lambda_c^{*}pD^{*}}$=3.75 for $J^p=1/2^{+}$.

For the $\Lambda_cpD$, $\Lambda_cpD^{*}$,
$KD_sD^{*}$, $\Sigma_cND$, and $\Sigma_cND^{*}$ vertices, we adopt the commonly employed Lagrangian densities as follows~\cite{Dong:2014ksa,Azevedo:2003qh,He:2016pfa},
\begin{align}
&{\cal{L}}_{\Lambda_cpD}=ig_{\Lambda_cpD}\bar{\Lambda}_c\gamma_5pD^{0}+{\rm H.c.},\\
&{\cal{L}}_{\Lambda_cpD^{*}}=g_{\Lambda_cpD^{*}}\bar{\Lambda}_c\gamma^{\mu}pD^{*0}_{\mu}+{\rm H.c.},\\
&{\cal{L}}_{KD_sD^{*}}=ig_{KD_sD^{*}}D^{*\mu}[\bar{D}_s\partial_{\mu}K-(\partial_{\mu}\bar{D}_s)K]+{\rm H.c.},\\
&{\cal{L}}_{\Sigma_cND}=-ig_{\Sigma_cND}\bar{N}\gamma_5\tau\cdot\Sigma_cD+{\rm H.c.},\\
&{\cal{L}}_{\Sigma_cND^{*}}=g_{\Sigma_cND^{*}}\bar{N}\gamma_{\mu}\tau\cdot\Sigma_cD^{*\mu}+{\rm H.c.}.
\end{align}
The coupling constants $g_{\Lambda_cpD}$=-13.98 and $g_{\Lambda_cpD^{*}}$=-5.20 are
determined from SU(4) invariant Lagrangians~\cite{Dong:2010xv} in terms of
$g_{\pi{}NN}$=13.45 and $g_{\rho{}NN}$=6.0. The coupling constants $g_{\Sigma_cND}=$2.69 and $g_{\Sigma_cND^{*}}=3.0$~\cite{Dong:2010xv}. The coupling constant
$g_{KD_sD^{*}}=5.0$ can also be evaluated from SU(4) symmetry~\cite{Azevedo:2003qh,Lin:1999ad,Oh:2000qr}.

When evaluating the scattering amplitude of the $K^{-}p\to{}D_s^{-}\Lambda_c^{*}$
reaction, we need to include the form factors because the hadrons are not pointlike
particles. We adopt here a common scheme
used in many previous works~\cite{He:2011jp,Xie:2015zga},
\begin{align}
F_{D^{*}}(q^2_{D^*},M_{ex})=\frac{\Lambda_{D^{*}}^2-M_{D^*}^2}{\Lambda_{D^{*}}^2-q_{D^*}^2},
\end{align}
for the $t$-channel $D^{*}$ meson exchange, and the form factor employed in Ref.~\cite{Shklyar:2005xg},
\begin{align}
F_B(q^2_{B},M_{B})=\frac{\Lambda_B^4}{\Lambda^4_B+(q_{B}^2-M_{B}^2)^2}.
\end{align}
for the exchanged baryon, $\Lambda^{*}_c$, $\Lambda_c(2286)^{+}$
or $\Sigma_c(2455)^{+}$. Here the $q_{(D^*,B)}$ and $M_{(D^*,B)}$ are the four-momentum and the mass of the exchanged $D^*$ meson (baryon),
respectively.  In this work, we use the cutoff parameters
$\Lambda_{D^*} = \Lambda_B =3$ GeV$^2$
for minimizing the free parameters. This values is chosen as the argument made in Refs.~\cite{Haidenbauer:2009ad,Dong:2014ksa}, and was employed in Refs.~\cite{He:2011jp,Xie:2015zga}. A variation of the cutoff will be made to show the sensitivity of the results on the cutoff.

\section{Kaon-induced  $\Lambda_c^{*}$  production with proton target}

  First, we will calculate the total cross section of the $K^{-}p\to{}\Lambda_c^{*}D_s$ reaction, which means the production possibility of the $\Lambda_c^*$.
By defining $s=(p_1+p_2)^2$, the corresponding
unpolarized differential cross section reads as
\begin{align}
\frac{d\sigma}{d\cos\theta}=\frac{M_pM_{\Lambda_c^{*}}}{16\pi{}s}\frac{|\vec{p}_{3cm}|}{|\vec{p}_{1cm}|}(\frac{1}{2}\sum_{s_c,s_2}|{\cal{M}}|^2),
\end{align}
where the $\theta$ is the scattering angle of the outgoing $D_s^{-}$ meson
relative to the beam direction, and $\vec{p}_{1cm}$ and $\vec{p}_{3cm}$
are the $K^{-}$ and $D_s^{-}$ three momenta in the center of mass frame.  The $M_p$
and $M_{\Lambda^{*}_c}$ are the masses of the proton and $\Lambda_c^{*}$,
respectively.

With the Lagrangians given in the previous section, the amplitude of the
$K^{-}(p_1)p(p_2)\to{}\Lambda_c^{*}(q)D_s(p_3)$ reaction can be botained as,
\begin{align}
{\cal{M}}^{1/2^{\pm}}&=g_{\Lambda_c^{*}pD^{*}}\bar{u}(q,s_c)\Gamma^{\mu\pm}u(p_2,s_2)G_{D^{*}}^{\mu\nu}(q_{D^{*}})\nonumber\\
                    &\times{}g_{KD_sD^{*}}(p_1^{\nu}+p_3^{\nu})F^2_{D^{*}}(q^2_{D^{*}},M_{D^{*}}),
\end{align}
where $\Gamma^{\mu\pm}=
\left( \gamma^{\mu} ,
    -\gamma_5\gamma^{\mu}
\right) $
, and the $\bar{u}(q,s_c)$ and $u(p_2,s_2)$ are the
Dirac spinors with $s_c$ ($q$) and $s_2$ ($p_2$) being the spins
(the four-momenta) of the outgoing $\Lambda_c^{*}$ and the initial
proton, respectively.

\begin{figure}[h!]
\begin{center}
\includegraphics[bb=20 400 900 750, clip,scale=0.45]{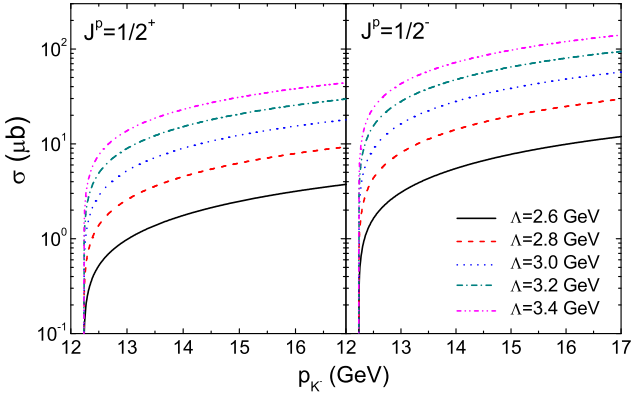}\\
\caption{(color online). The total cross section $\sigma$ for the $K^{-}p\to{}D_s^{-}
\Lambda_c^{*}$ reaction as a function of the beam momentum $p_{K^{-}}$
for the cases of the $\Lambda_c^*$ with $J^{p}=1/2^+$ (left panel) and $1/2^-$ (right panel).} \label{cross22}
\end{center}
\end{figure}

The numerical results about total cross section of the $K^{-}p\to{}D_s^{-}\Lambda_c^{*}$ reaction is presented in Fig.~\ref{cross22}. Because the cutoff can not be well determined, the results at cutoffs deviated from 3 GeV are also presented.

The results show that the total cross section increases sharply near the $D^-_s\Lambda_c^*$ threshold. At higher energies, the cross section increases continuously but relatively slowly compared with near threshold. With the increase of the cutoff, the total cross section will increase, and generally speaking, the total cross for spin-parity assignment $1/2^+$ is smaller than these for $1/2^-$.
At a beam momentum about 15 GeV the order of magnitude  are about 10 $\mu$b, which is considerably large for the experimentally observation of the $\Lambda^*_c$ with the current experimental technology.

\section{The $K^{-}p\to D_s^{-}D^0p$ reaction}

Since the $\Lambda_c^*$ can not be observed directly, the $D^0p$ channel of the $\Lambda^*_c$ decay, which is the observation channel of the $\Lambda^*_c$ at \babar, will be introduced to give a more realistic prediction for the observation of the $\Lambda_c^*$ in experiment. Here we consider the subsequential decay of the $\Lambda^*_c$ after produced in the $K^-p\to\Lambda^*_cD_s^-$ reaction, i.e. the $K^{-}p\to{}D_s^{-}D^{0}p$
reaction, which is illustrated in Fig.~\ref{feydiagrams23}. The contributions from the $\Lambda_c(2286)$ and the $\Sigma_c(2455)$ will be included as background.
It is a two to three body process, the cross
section can be obtained from the amplitude as,
\begin{align}
&d\sigma(K^{-}p\to{}D_s^{-}D^{0}p)\nonumber\\
&=\frac{M_N}{2\sqrt{(p_1\cdot{p_2})^2-M^2_{K^{-}}M_N^2}}\sum_{s_i,s_f}|{\cal{M}}({K^{-}p\to{}D_s^{-}D^{0}p})|^2\nonumber\\
&\times\frac{d^3\vec{p}_3}{2E_3}\frac{d^3\vec{p}_4}{2E_4}\frac{M_Nd^3\vec{p}_5}{E_5}\delta^4(p_1+p_2-p_3-p_4-p_5),\label{cs23}
\end{align}
where $E_3$,~$E_4$ and $E_5$ stand for energy of $D_s^{-}$,~$D^-$ and final proton, respectively.
And, the $M_{K^{-}}$ stand for mass of beam particle.

The amplitude ${\cal{M}}(K^{-}p\to{}D_s^{-}D^{0}p)$ can be obtained with Lagrangians given in Sec.~II as,
\begin{align}
{\cal{M}}^{1/2^{+}}&(K^{-}p\to{}D_s^{-}D^{0}p)=i\frac{g_{\Lambda^{*}_cpD}g_{\Lambda_c^{*}pD^{*}}g_{KD_sD^{*}}}{q^2-M^2_{\Lambda^{*}_c}+iM_{\Lambda_c^{*}}\Gamma_{\Lambda^{*}_c}}\nonumber\\
                 &\times{}\frac{1}{k^2-M^2_{D^{*0}}}F_{\Lambda_c^{*}}(q^2,M_{\Lambda_c^{*}})F^2_{D^{*}}(k^2,M_{D^{*}})(p_{1\mu}+p_{3\mu})\nonumber\\
                     &\times{}\bar{u}(p_5,s_5)\gamma_5(q\!\!\!/+M_{\Lambda_c^{*}})(\gamma^{\mu}-\frac{k\!\!\!/k^{\mu}}{M^2_{D^{*}}})u(p_2,s_2),\\
{\cal{M}}^{1/2^{-}}&({K^{-}p\to{}D_s^{-}D^{0}p})=\frac{f_{\Lambda^{*}_cpD}f_{\Lambda_c^{*}pD^{*}}g_{KD_sD^{*}}}{q^2-M^2_{\Lambda^{*}_c}+iM_{\Lambda_c^{*}}\Gamma_{\Lambda^{*}_c}}\nonumber\\
                   &\times{}\frac{1}{k^2-M^2_{D^{*0}}}F_{\Lambda_c^{*}}(q^2,M_{\Lambda_c^{*}})F^2_{D^{*}}(k^2,M_{D^{*}})(p_{1\mu}+p_{3\mu})\nonumber\\
                   &\times{}\bar{u}(p_5,s_5)(q\!\!\!/+M_{\Lambda_c^{*}})\gamma_5(\gamma^{\mu}-\frac{k\!\!\!/k^{\mu}}{M^2_{D^{*}}})u(p_2,s_2),
\end{align}
and the amplitudes for the $\Lambda_c(2286)$ and the $\Sigma_c(2455)$ can be obtained analogously.
Here we take $\Gamma=0$ MeV for the $\Lambda_{c}(2286)$ and the $\Sigma_c(2455)$ states because of their very small values of the experimental decay width. For the $\Lambda_c^{*}$
resonance which we focus on, a value as $\Gamma=17$ MeV is adopted according to decay width observed at \babar~which  is consistent with the one observed at Belle~\cite{K. A. Olive}.

\begin{figure}[h!]
\begin{center}
\includegraphics[bb=18 400 900 755, clip,scale=0.44]{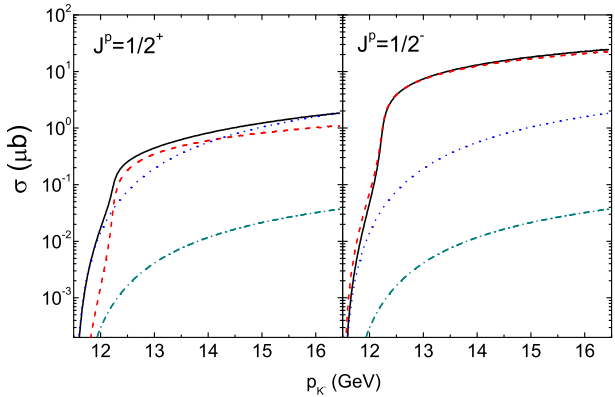}
\caption{(color online). Total cross section $\sigma$ for the $K^{-}p\to{}D_s^{-}D^{0}p$
reaction as a function of the beam momentum $p_{K^{-}}$ for the $\Lambda_c^{*}$ with $J^{p}=1/2^+$ (left panel)
and $J^{p}=1/2^-$ (right panel).  The dashed, dotted, and dash-dotted
curves stand for the contributions from the $\Lambda_c^{*}$, the  $\Lambda_c(2286)$,
and the $\Sigma_c(2455)$, respectively.  The total contribution is shown by
the solid line.} \label{cross-plus23}
\end{center}
\end{figure}

With the formalism and ingredients given above, the cross section against
the beam momentum $p_{K^{-}}$ for the $K^{-}p\to{}D_s^{-}D^{0}p$
reaction is calculated by using a Monte Carlo multiparticle phase
space integration program, FOWL program, and checked with a direct integration with Eq.~(\ref{cs23}).  The theoretical results at a cutoff $\Lambda=3.0$ GeV
for the beam momentum $p_{K^{-}}$ from near threshold upto 16.5 GeV are shown
in Fig.~\ref{cross-plus23}.  The ontributions from the ground $\Lambda_c(2286)$ state
and the $\Lambda_c(2455)$ state are also presented in the same figure.

As in the  $\Lambda_c^*$ production, the total cross section for an assignment of spin parity of the $\Lambda_c^*$ as $J^P=1/2^-$ is much larger than that for assignment as $1/2^+$. The order of magnitude of total cross section is about 1 and 10 $\mu$b for positive and negative parity, respectively. The total cross section of the background contribution from the $\Sigma_c(2455)$ is much smaller than other contributions. The contribution from the $\Lambda^*_c$ which we focus on in this work is much larger than other contributions if the spin parity of the $\Lambda^*_c$ is chosen as $1/2^-$, which makes the observation of the $\Lambda_c^*$ become easy to do in experiment. However, if the $\Lambda_c^*$ carries a spin parity of $1/2^+$, the total cross section from the $\Lambda^*_c$ is much smaller and comparable with the background contribution especially that from the $\Lambda^*_c(2286)$.

\begin{figure}[h!]
\begin{center}
\includegraphics[bb=10 36 510 690, clip,scale=0.44]{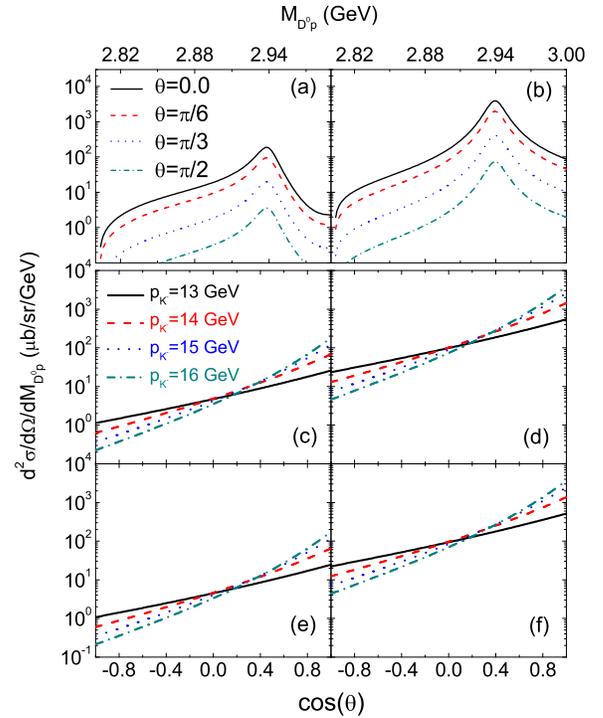}
\caption{(color online).  The second order differential cross section $d^2\sigma/d\Omega/dM_{D^0p}$ for the $K^{-}p\to{}D_s^{-}D^{0}p$
reaction as a function of invariant mass of the final $pD^{0}$ two-body system $M_{D^0p}$ (panels a and b), and the scattering angle $\cos\theta$ (panels c, d, e, and f). The results for the spin parity $J^P=1/2^+$ of the $\Lambda^*_c$ are presented in the panels (a), (c), and (e) , and these for $1/2^+$ in the panels (b), (d), and (f). The panels (c) and (d) are for the differential cross sections with background, and the panels (e) and (f) for these without background.} \label{dftwo}
\end{center}
\end{figure}

To give more theoretical information about the $\Lambda_c^*$ production in the $K^{-}p\to{}D_s^{-}D^{0}p$ reaction, we present the second order differential
cross section $d^2\sigma/d\Omega/dM_{D^0p}$ as a function of invariant mass of the final $pD^{0}$ two-body system at a momentum of kaon beam as $p_{K^-}=16$ GeV in Fig.~\ref{dftwo}(a) and Fig.~\ref{dftwo}(b) for positive and negative parities of the $\Lambda^*_c$, respectively.

The results at typical scattering angles, $\theta$=0, $\pi/6$, $\pi/3$, and $\pi/2$ are given. An obvious peak can be found around the invariant mass $M_{D^0p}=2.94$ GeV as expected. The differential cross section is largest at extreme forward angle  and decrease with the increase of the scattering angle. It can be seen more clearly at the differential cross sections as a function of
scattering angle at invariant mass $M_{pD^{0}}=2.94$ GeV in Fig.~\ref{dftwo}(c-f). In Fig.~\ref{dftwo}(c) and Fig.~\ref{dftwo}(d),  the results for the $K^{-}p\to{}D_s^{-}D^{0}p$ reaction at different momenta of kaon beam $p_{K^-}$ are presented. The results show that  with the increase of the kaon momentum, the increase of the total cross section as shown in FIg.~\ref{cross-plus23} is mainly from the increase of the differential cross section at forward angles. We also present the results of for the $K^{-}p\to{}D_s^{-}D^{0}p$ reaction through $\Lambda^*_c$ only in Fig.~\ref{dftwo}(e) and Fig.~\ref{dftwo}(f), which suggest the effects of the background on the the differential cross sections as a function of
scattering angle is very small.  With the increase of the momenta of the kaon beam, more of the $\Lambda(2940)$ are produced at forward angles. The results suggest that it is better to observe the $\Lambda^*_c$ at forward angles especially at high energies.

\begin{figure}[h!]
\begin{center}
\includegraphics[bb=58 330 900 750, clip,scale=0.48]{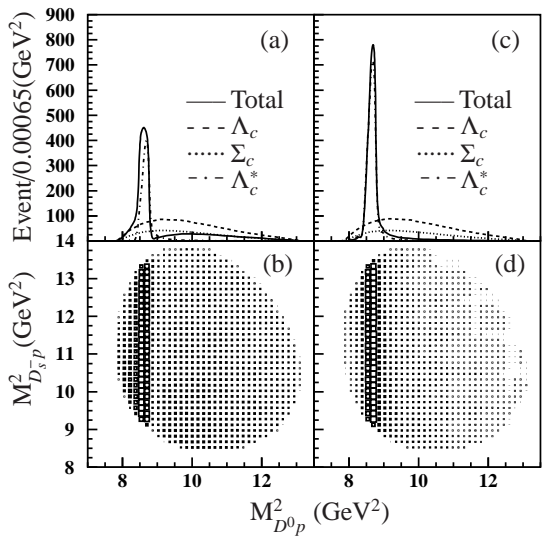}
\caption{(color online).  The invariant mass distribution (panels a and c) and Dalitz Plot (panels b and d) for the $K^{-}p\to{}D_s^{-}D^{0}p$
reaction at beam energy $p_{K^{-}}=16$ GeV. The panels (a) and (b) are for the case with spin parity $J^{p}=1/2^{+}$  of the $\Lambda^*_c$ , and panels (c) and (d) for the case with $J^{p}=1/2^{-}$.} \label{dalt}
\end{center}
\end{figure}

In addition, we present
the invariant mass distribution and of the Dalitz plot the
$K^{-}p\to{}D_{s}^{-}D^{0}p$ reaction in Fig.~\ref{dalt}.  It is interesting to see that the peaks are obvious and at invariant mass  $M_{pD^{0}}=2.94$ GeV  with both assignments of the spin parity of the $\Lambda^*_c$. For the assignment $1/2^+$ of the $\Lambda^*_c$, though the  background  provides considerable contribution to the total cross section as shown in Fig.~\ref{cross-plus23}(a), its contribution to the invariant mass spectrum change slowly because both the $\Lambda_c(2455)$ and the $\Sigma_c (2286)$ are below the $D^0p$ threshold. The $\Lambda^*_c$ exhibits itself as a sharp peak near 2.94 GeV, which makes it easy to observe in the experiment with both positive and negative parities shown in Fig.~\ref{dalt}(a) and Fig.~\ref{dalt}(c), respectively.  In Fig.~\ref{dalt}(b) and Fig.~\ref{dalt}(d) the Dalitz plots of the $K^{-}p\to{}D_{s}^{-}D^{0}p$ reaction are presented. With both assignments of the spin parities of the $\Lambda^*_c$, an obvious band for the $\Lambda^*_c$ can be found near $M_{D^0p}=2.94$ GeV.

\section{SUMMARY}
In this work, we perform a calculation of the $\Lambda_c^{*}$ production
in the $K^{-}p\to{}D^{-}_s\Lambda_c^{*}$ and the $K^{-}p\to{}D_s^{-}D^{0}p$
reactions within the effective Lagrangian approach to study the possibility to study the charmed baryon $\Lambda^*_c$ with kaon beam on a proton target in experiment. The total cross section of the $\Lambda^*_c$ productions is found at an order of magnitude about 10 $\mu$b. After considering the subsequential decay of the $\Lambda^*_c$ in $D^0p$ channel, the  results shown that the signal of the $\Lambda^*_c$ is significant in the $D^0p$ invariant mass spectrum and the Dalitz plot of the $K^{-}p\to{}D_s^{-}D^{0}p$
reaction. The large production possibility and the significant of the signal of the $\Lambda_c^*$ in the kaon-induced production on a proton target suggest that a study of the $\Lambda_c(2940)$ with kaon beam  are promising in  further experiment.

\section*{Acknowledgments}
This project was partially supported by the Major State Basic
Research Development Program in China (No. 2014CB845400), the
National Natural Science Foundation of China (Grants No. 11475227,
No. 11275235, No. 11035006, No.11175220, and No.11675228) and the Chinese Academy of
Sciences (the Knowledge Innovation Project under Grant No.
KJCX2-EW-N01, Century program under Grant No. Y101020BR0). It is
also supported by the Open Project Program of State Key Laboratory
of Theoretical Physics, Institute of Theoretical Physics, Chinese
Academy of Sciences, China (No. Y5KF151CJ1).

\end{document}